\documentclass[journal,twoside,web]{ieeecolor}
\usepackage{generic}
\usepackage{cite}
\usepackage{amsmath,amssymb,amsfonts}
\usepackage{graphicx}

\usepackage{siunitx}
\usepackage{wrapfig}
\usepackage[colorlinks=true, allcolors=blue]{hyperref}
\usepackage[capitalise]{cleveref}
\usepackage{subcaption}
\usepackage{booktabs}
\usepackage{float}
\usepackage[normalem]{ulem}
\DeclareSIUnit\dec{dec}

\newcommand{\mos}{\rm{MoS$_2$}}
\newcommand{\nbs}{\rm{NbS$_2$}}

\begin{document}

\title{Two-dimensional field-effect transistors based on lateral heterojunctions of transition metal dichalcogenides: dissipative quantum transport modeling}
\author{Giuseppe Lovarelli, Fabrizio Mazziotti, Demetrio Logoteta and Giuseppe Iannaccone, \IEEEmembership{Fellow, IEEE}
\thanks{This work was partially supported by the European Union Horizon 2020 Framework Programme under the “Quantum Engineering for Machine Learning” (QUEFORMAL) project (Grant Agreement No. 829035), by the by the Italian MIUR through the PRIN project FIVE2D and by the Italian Ministry of University and Research through the ForeLab Departments of Excellence Grant.}
\thanks{Giuseppe Lovarelli was with Dipartimento di Fisica, Università di Pisa, Largo Bruno Pontecorvo 3, 56127 Pisa, Italy and with Dipartimento di Ingegneria dell’Informazione, Università di Pisa, Via Girolamo Caruso 16, 56122 Pisa, Italy (e-mail: giuseppe.lovarelli@phd.unipi.it).}
\thanks{Fabrizio Mazziotti is with Dipartimento di Ingegneria dell’Informazione, Università di Pisa, Via Girolamo Caruso 16, 56122 Pisa, Italy (e-mail: f.mazziotti1@phd.unipi.it).}
\thanks{Demetrio Logoteta is with Dipartimento di Ingegneria dell'Informazione, Elettronica e Telecomunicazioni, Università di Roma La Sapienza, via Eudossiana  18, 00184, Roma, Italy, (e-mail: demetrio.logoteta@uniroma1.it).}
\thanks{Giuseppe Iannaccone is with Dipartimento di Ingegneria dell’Informazione, Università di Pisa, Via Girolamo Caruso 16, 56122 Pisa, Italy (e-mail: giuseppe.iannaccone@unipi.it).}}

\maketitle

\begin{abstract}
Reducing the contact resistance of field-effect transistors based on two-dimensional materials is one of the key improvements required to enable the integration of such transistors in an advanced semiconductor manufacturing process. Suitably designed lateral heterojunctions provide an opportunity to independently tailor the contact and channel properties in order to optimize contact resistance. Inspired by the recent experimental demonstration of a two-dimensional $p$-type Schottky barrier, here we use quantum transport simulations to estimate the performance of \textit{p}-type transistors in which the channel consists of a lateral heterostructure of {\nbs/\mos/\nbs} (semimetal-semiconductor-semimetal). We find that the gate alignment with the channel is a critical design parameter,  strongly influencing the capability of the gate to modulate the Schottky barrier at the {\mos/\nbs} interface. This effect is also found to significantly affect the scaling behavior of the device.
\end{abstract}


\section{Introduction}
\label{sec:introduction}
\IEEEPARstart{T}{he} possibility to use two-dimensional (2D) materials in an industrial complementary metal-oxide semiconductor (CMOS) process depends several factors, but few are as critical as the capability to fabricate low-resistance metal contacts. 
Unfortunately, 2D semiconductors tend to form high Schottky barriers with bulk metals, and the difficulty in achieving high doping levels in monolayers hinders the adoption of the conventional approach based on thinning these barriers enough to make them transparent to carrier tunnelling~\cite{tersoff1984,yang2014,allain2015}.

Contacting semiconducting monolayers by defining vertical or lateral heterojunctions with 2D metals or quasi-metals has attracted significant interest~\cite{jena2014,kaushik2016,xie2017,schulman2018,cai2022,chen2022,choi2022,schneider2021}, especially in the perspective of developing all-2D circuits where all circuit components, including metallizations and insulating regions, are based on 2D materials.

In vertical heterojunctions, monolayers are weakly coupled by van der Waals interactions.
This prevents the undesired occurrence of interfacial spurious states and Fermi level pinning~\cite{tersoff1984,kim2017,wang2019,liu2022}; however, the small coupling also entails low carrier injection~\cite{wang2019}. 
On the other hand, the edge contact between monolayers in lateral heterojunctions is characterized by a stronger coupling due to the covalent character of the atomic bonds.
Provided that the quality of the interface is controlled and the height of the Schottky barrier is low enough, this translates to potentially efficient carrier injection~\cite{zhang2018,shen2021}.

The capability to engineer both $n$- and $p$-type contacts is another essential requirement to implement a CMOS technology.
Unfortunately, realizing $p$-type field-effect transistors (FETs) based on 2D materials has proven extremely challenging, due to the strong influence of the commonly used dielectric substrates, that tend to induce an $n$-type doping in the channel and of Fermi level pinning at the conduction band~\cite{he2019}.
In this respect, recent experiments have demonstrated the fabrication of $p$-type Schottky diodes based on a lateral heterojunction of {\nbs} and {\mos} in their 2H-form~\cite{wang2023}.
The reported, significant, current modulation indicates an effective reduction of the parasitic resistance associated to the {\mos} contact when the {\nbs} buffer region is included.

In this paper we study the physics and the  performance perspectives of a $p$-type FET based on a monolayer {\mos} channel connected to monolayer {\nbs} access regions through lateral heterojunctions, using quantum transport simulations and critical materials parameters calibrated with experiments as discussed in Ref.~\cite{wang2023}.
Our results indicate that the phonon-assisted interband tunneling at the {\nbs/\mos} interface plays a major role in determining device behavior.
It also translates into a pronounced sensitivity of the device to some architectural details, such as the alignment between the gate electrode and the channel, that should be carefully addressed in order to optimize  performance.

We find that the 2D heterojunction FET is not competitive in terms of performance with silicon FETs for front-end devices, as can be observed by considering the IRDS (International Roadmap for Devices and Systems)~\cite{IRDS2021}, but it has very interesting performance in terms of devices suitable for 3D integration (i.e., the vertical stacking of planar layers of transistors), given that it can be fabricated with back-end-of-the-line (BEOL) processing~\cite{Iannaccone2018}.

The paper is organized as follows.
In Section II, we present our model and our simulation approach.
In Section III, we discuss our results in terms of the device physics, performance and optimization opportunities.
Our concluding remarks are drawn in Section IV.

\section{Model}

Following Ref.~\cite{wang2023}, we model the isolated monolayers by means of the two-bands first-nearest neighbor tight-binding Hamiltonian
\begin{equation}
\label{HAMILT}
    H_{2D,j} \left( \textbf{k} \right) = 
    \begin{pmatrix}
    E_{V,j} & t_j \, f_{2D} \left( \textbf{k} \right) \\
    t^*_j \, f_{2D}^* \left( \textbf{k} \right) & E_{C,j}
    \end{pmatrix} \, ,
\end{equation}
where $E_{C,j}$ and $E_{V,j}$ are the conduction band minimum and valence band maximum, respectively, $t_j$ is the hopping parameter, with $j=$ ({\mos}, {\nbs}). 
In addition, 
\begin{equation}
    f_{2D} \left( \textbf{k} \right) = 1 + 2\, {\rm cos} \left( \frac a2 k_{x} \right) e^{i \frac{\sqrt{3}}{2} k_{y}}
\end{equation}
is the Bloch function that describes the periodicity of the hexagonal lattice with lattice constant $a$.
Due to the relatively lower {\nbs} in-plane stiffness with respect to {\mos}~\cite{kang2015,Sun2018}, the channel was assumed unstrained, while the source and drain extensions were assumed compressively strained in the transverse direction. Accordingly, the lattice constant of {\nbs} was assumed the same as that of {\mos}.

In agreement with the experimental and theoretical analyses in~\cite{wang2023}, the {\mos/\nbs} Schottky barrier height, i.e., the difference between $\mu_{\textrm{NbS}_2}$, the electrochemical potential of {\nbs}, and the top of the {\mos} valence band is $\SI{0.17}{\eV}$. 
Furthermore, the degeneracy of the {\nbs} conduction band is 
$\mu_{\textrm{NbS}_2} -  E_{C,\textrm{NbS}_2} = \SI{0.34}{\eV}$. For the {\nbs}, the values of $E_C$, $E_V$ and $t$ were chosen in order to reproduce the density of states near the Fermi level $E_{F}$, while for the {\mos}, we chose these parameters in order to reproduce the valence band curvature at the $K$-point.

\begin{figure}
    \centering
    \includegraphics[width=.9\columnwidth]{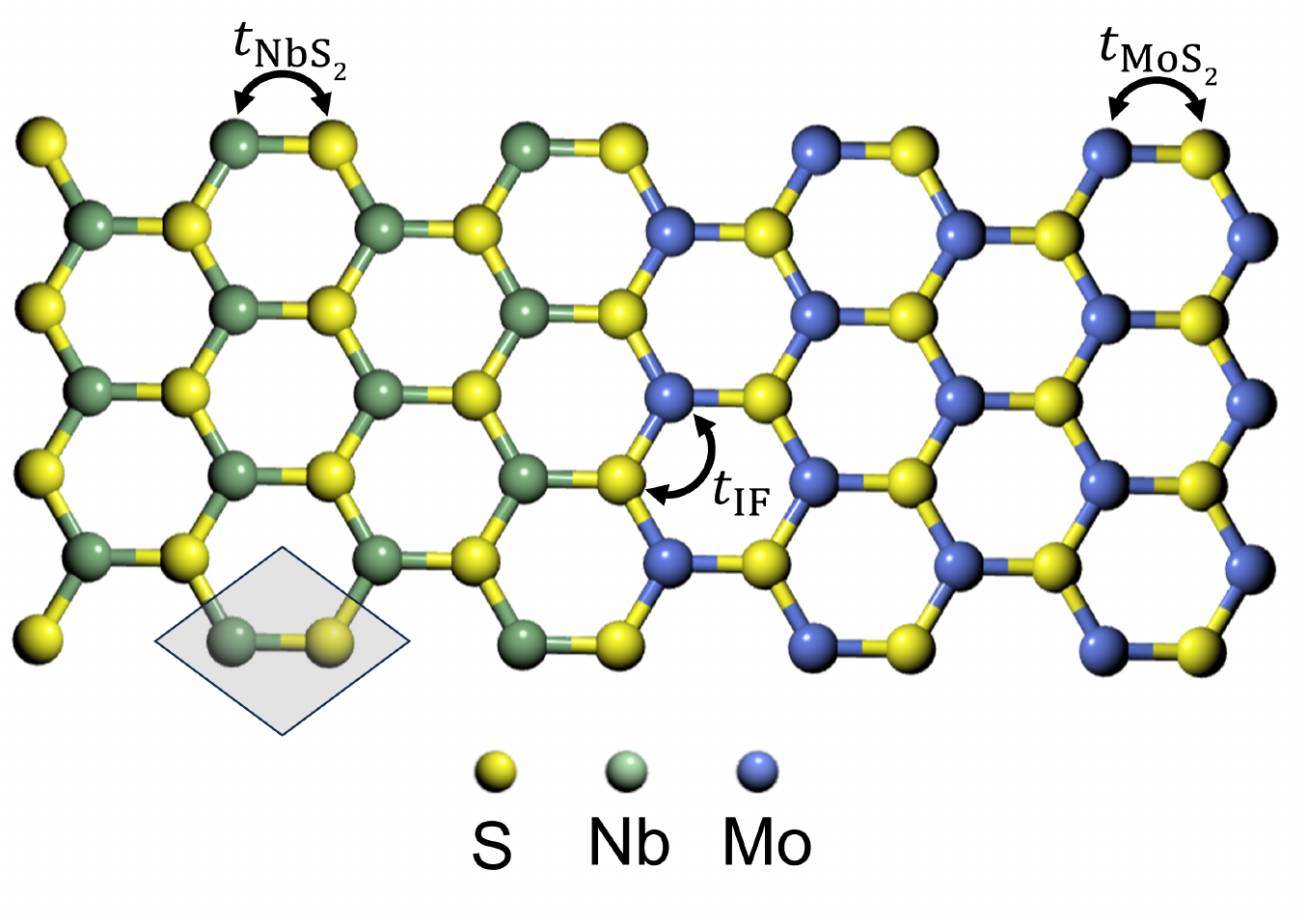}
    \caption{Sketch of the {\mos}/{\nbs} heterointerface. The shaded region depicts the unit cell of the lattice. The tight-binding parameters are also indicated.
    }
    \label{fig:lh_junction}
\end{figure}

As pointed out in~\cite{wang2023}, the exact value of the lattice constant of Nb-doped {\mos} has a negligible influence on the band alignment with {\nbs}. Therefore, as a simplifying approximation, we considered it equal to the lattice constant of {\nbs}.
The interface between the monolayers is assumed defectless and, according to the above-mentioned approximation, unstrained (see~\cref{fig:lh_junction}).
As a zeroth-order approximation, the value of the hopping parameter between Mo and S atoms at the interface was assumed the same as in the free-standing monolayer. In practice, it can both increase or decrease with respect to this value, but for small enough deviations the transmission through the interface will remain dominated by the Schottky barrier.

The hopping parameter $t_{\rm IF}$ between {\nbs} sulphur atoms and {\mos} molybdenum atoms is assumed equal to $t_{\textrm{NbS}_2}$: 
due to the closeness between the values of $t_{\textrm{MoS}_2}$ and $t_{\textrm{NbS}_2}$, no significant differences are expected by setting $t_{\rm IF}$ to $t_{\textrm{NbS}_2}$ or to an intermediate value between $t_{\textrm{MoS}_2}$ and $t_{\textrm{NbS}_2}$.
We report the parameters of the model in~\cref{tab:parametri}.

Transport simulations were performed within the non-equilibrium Green's function formalism~\cite{DATTA2000microsuper} by using an in-house code~\cite{Cao2016,Logoteta2020}.
An orthorhombic elementary cell is designated to describe the system in real space in the direction of transport, and the device is assumed periodic in the lateral direction orthogonal to it.
To sample the Brillouin zone along this transverse direction, a set of 100 transverse wave vectors were taken into account in simulations.   
In order to obtain a self-consistent solution, the transport equations were nonlinearly coupled with the 2D Poisson equation in the device cross section.
The resulting system of equations was iteratively solved by adopting a fixed-point approach.
Convergence was considered achieved when the maximum difference between the potential energy profile in consecutive iterations was less than 1 meV. 

Electron-phonon scattering was included in the transport equations within the self-consistent Born approximation through diagonal self-energies~\cite{SCBA} and by assuming an elastic and dispersionless approximation for acoustic and optical phonons, respectively.
The optical phonon energy was obtained as the arithmetical average of the longitudinal, transverse and homopolar optical branches in MoS$_2$, and set to $\hbar\bar{\omega}=50$ meV~\cite{Xiaodong2013}.
Hole-phonon coupling was described within the deformation potential model.
According to our long-wavelength approximation ~\cite{Lundstrom2000} and taking into account the suppression of the hole intervalley scattering in transition metal dichalcogenides due to the spin-valley locking effect~\cite{Schaibley2016}, only the intravalley hole scattering at the $K$ point was considered~\cite{Zhenghe2014}.

The values of the acoustic and optical deformation potentials ($D_{\rm op}$ and $D_{\rm ac}$, respectively) are reported in~\cref{tab:parametri}.

\begin{table}
    \centering
    \caption{\label{tab:parametri}Tight-binding parameters and deformation potentials used in simulations}
    \begin{tabular}{c c}
    \toprule
        $E_{\rm V,\textrm{NbS}_2}$  & $-\SI{5.59}{\eV}$ \\
        $E_{\rm C,\textrm{NbS}_2}$  & $-\SI{0.34}{\eV}$ \\
        $E_{\rm V,\textrm{MoS}_2}$  & $-\SI{0.17}{\eV}$ \\
        $E_{\rm C,\textrm{MoS}_2}$  & \phantom{$-$}$\SI{1.93}{\eV}$ \\
        $t_{\textrm{NbS}_2}$                 & \SI{-0.91}{\eV} \\
        $t_{\textrm{MoS}_2}$                 & \SI{-1.20}{\eV} \\
        \hline
        $\hbar \bar{\omega}_{\rm ph}$ & \phantom{$-$}\SI{0.05}{\eV} \\
        $D_{\rm op}$                & \SI{4.6e8}{\eV/ \centi\meter}~\cite{Zhenghe2014} \\
        $D_{\rm ac}$                & \phantom{$-$}\SI{2.5}{\eV}~\cite{Zhenghe2014} \\
    \bottomrule
    \end{tabular}
\end{table}

\begin{figure}
    \centering
    \includegraphics{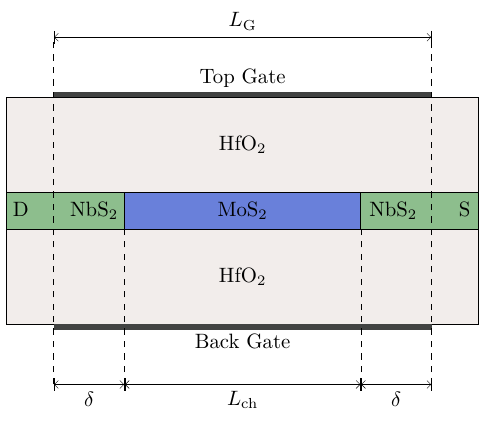}
    \caption{Sketch of the cross section of the {\nbs/\mos/\nbs} transistor.
    }
    \label{fig:dispositivo}
\end{figure}

\section{Results and discussion}
\begin{figure*}
	\centering
	\begin{subfigure}[t]{0.33\textwidth}
		\centering
		\includegraphics[height=4.8cm]{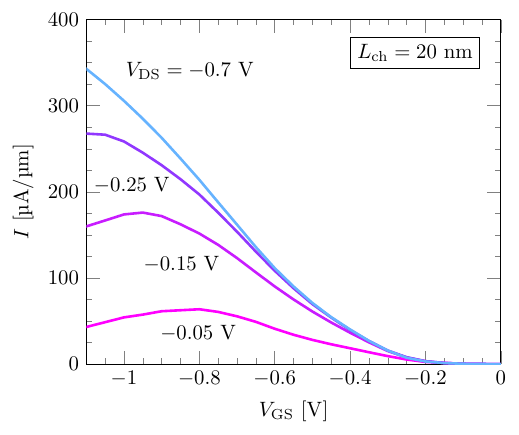}
		\caption{ }
		\label{fig:ivgs_20nm}
	\end{subfigure}\hfil
	\begin{subfigure}[t]{0.67\textwidth}
		\centering
		\includegraphics[height=4.8cm]{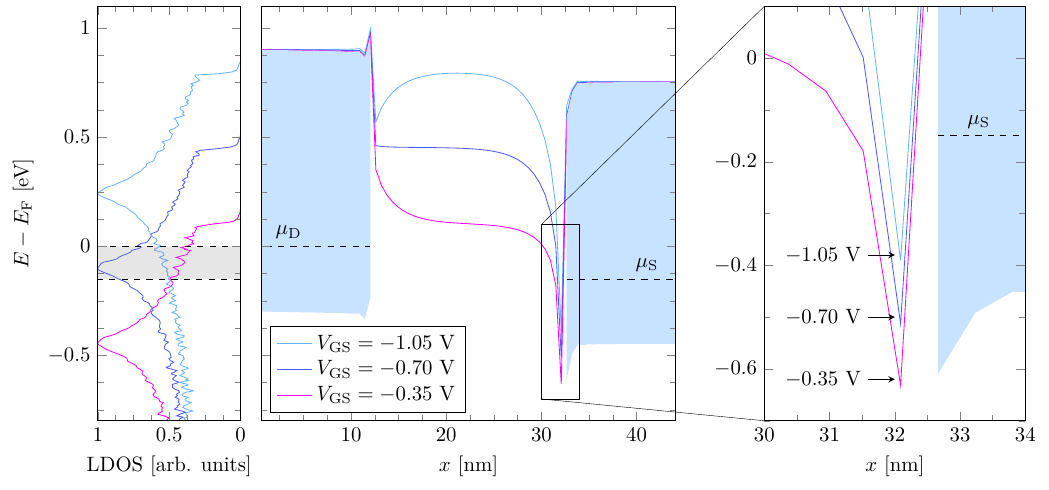}
		\caption{ }
		\label{fig:dos_bande_barriera_20nm}
	\end{subfigure}\\
	\begin{subfigure}[t]{0.32\textwidth}
		\centering
		\includegraphics[height=4.8cm]{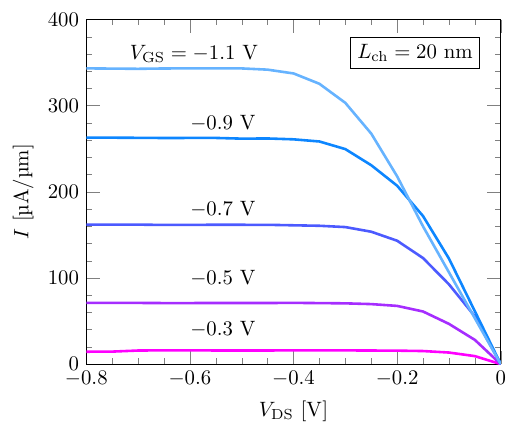}
		\caption{ }
		\label{fig:ivds_20nm}
	\end{subfigure}
	\begin{subfigure}[t]{0.32\textwidth}
		\centering
		\includegraphics[height=4.6cm]{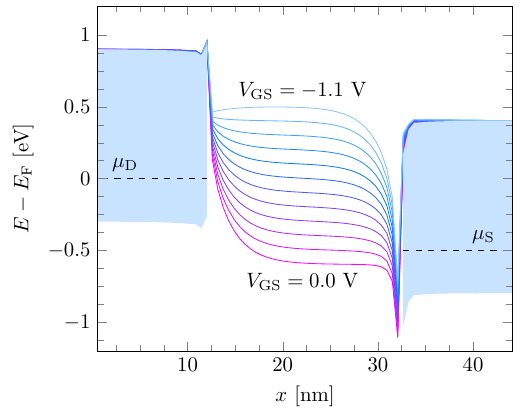}
		\caption{ }
		\label{fig:bande_20nm}
	\end{subfigure}
	\begin{subfigure}[t]{0.32\textwidth}
		\centering
		\includegraphics[height=4.8cm]{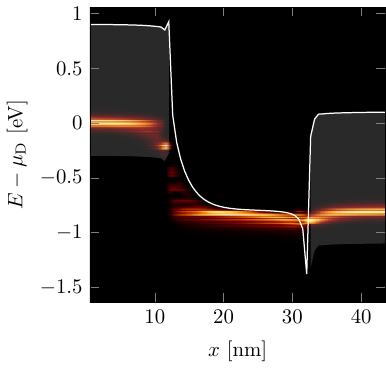}
		\caption{ }
		\label{fig:jdens_20nm}
	\end{subfigure}\\
	\caption{(a) Transistor transfer characteristics for different values of $V_{\rm DS}$. (b) Band diagram along the device (center panel) and local density of states in the center of the channel (left panel) at $V_{\rm DS}=-0.15$ V and for three $V_{\rm GS}$ values close to the current peak. The blue-colored regions cover the energy window of the {\nbs} band involved in transport. The right panel is an enlargement of the band diagram close to the channel/source interface. (c) Transistor output characteristics. (d) Band diagram along the device for several values of $V_{\rm GS}$. (e) Current spectrum at $V_{\rm DS}=-0.8 \ \si{\volt}$ and $V_{\rm GS}=0.1 \ \si{\volt}$. For all the panels, $L_{\rm ch}=20 \ \si{\nano\meter}$.
	}
	\label{fig:panel_20nm}
\end{figure*}

The cross-section of the {\nbs/\mos/\nbs} transistor is sketched in~\cref{fig:dispositivo}.
The current flow is controlled by means of a top and a bottom gate, separated from the channel by a $\SI{3.4}{\nano\meter}$-thick HfO$_2$ layer with relative permittivity $\epsilon_r = 20$.
The gate length $L_{\rm G}$ can be larger or smaller than the {\mos} channel length $L_{\rm ch}$; this difference is quantified by the algebraic parameter $\delta=(L_{\rm G}-L_{\rm ch})/2$.
The part of the {\nbs} source (S) and drain (D) extensions included in the domain of the Poisson equation are $\SI{12}{\nano\meter}$ long. Further enlarging these regions does not modify the simulation results.
According to the estimates in Ref.~\cite{wang2023}, the {\mos} channel is \textit{p}-doped at a concentration $\SI{6e12}{\centi\meter^{-2}}$.

\cref{fig:ivgs_20nm} illustrates the transfer characteristics of the transistor for $L_{\rm ch}=20$ nm, $\delta=0$ and different values of $V_{\rm DS}$.
A pronounced maximum of the current is observed, that shifts towards more negative $V_{\rm GS}$ as $|V_{\rm DS}|$ increases.
This effect can be explained with the help of~\cref{fig:dos_bande_barriera_20nm}, that reports the band diagram of the transistor (central panel) for $V_{\rm DS}=-0.15$ V and three values of $V_{\rm GS}$, located before, at, and after the current peak.
The corresponding local density of states (LDOS) in the middle of the channel is shown in the left panel of the figure, while the right panel presents a magnified view of the band diagram close to the {\mos/\nbs} interface at the source.
The LDOS exhibits a peak, corresponding to the local flattening of the {\mos} valence band at the $M$ point.
The current is mainly limited by two factors: the density of available states in the channel between the source and drain electrochemical potentials (grey stripe in~\cref{fig:dos_bande_barriera_20nm}), and the width of the tunneling barrier at the source-channel interface.
As the gate voltage is swept through the current peak, the barrier thins, while the density of states decreases.
The competition between the two phenomena induces the current peak. 

When $\left| V_{\rm DS} \right|$ increases, the tunneling barrier at the source/channel interface at a given energy widens.
As a consequence, more negative gate voltages are needed to reach the current peak. 

At $V_{\rm DS}=-0.8$ V and $-0.6$ V, the $I_{\rm ON}/I_{\rm OFF}$ ratio equals 2330 and 1290, respectively.
These value are obtained by setting $I_{\rm OFF} = \SI{0.1}{\micro\ampere/\micro\meter}$~\cite{IRDS2021} and computing $I_{\rm ON}$ at $V_{\rm GS} = V_{\rm GS, OFF} + V_{\rm DS}$, where $V_{\rm GS,  OFF}$ is the gate voltage at which $I_{\rm D} = I_{\rm OFF}$.
We maintain these definitions through the paper.

The output characteristics of the transistor with $L_{\rm ch}=20$ nm are reported in~\cref{fig:ivds_20nm}. 
An excellent current saturation at large bias is observed.
Furthermore, we can observe that at low $|V_{\rm DS}|$ the curve at $V_{\rm GS}=-1.1$ V intersects with the curve at lower $V_{\rm GS}$. 
This is simply due to the presence of the peak in the $I-V_{\rm GS}$ characteristics shown in Figure \ref{fig:dos_bande_barriera_20nm}. 
\cref{fig:bande_20nm} shows the band diagram along the device for several values of $V_{\rm GS}$, which confirms an almost perfect electrostatic control of the gate over the channel potential, corresponding to a gate efficiency $\partial E_{\rm V}/\partial V_{\rm GS}\simeq 0.98$, where $E_{\rm V}$ is the top of the valence band at the center of the channel. 

Despite this almost ideal electrostatic control of the gate over the channel, the subthreshold swing of the device is about $95 \ \si{\milli\volt/\dec}$ and is thus degraded with respect to the thermionic limit of $60 \ \si{\milli\volt/\dec}$.
This is a direct consequence of the carrier energy relaxation due to phonon emission, which allows interband electron transmission at the channel-drain interface also when the the top of the valence band in the middle of the channel is pushed below the lower edge of the {\nbs} band.
This effect is illustrated in~\cref{fig:jdens_20nm}, which reports the current spectral density when the device operates in the subthreshold region.

\begin{figure*}
	\centering
	\begin{subfigure}[t]{0.2\textwidth}
		\centering
		\includegraphics[height=4.6cm]{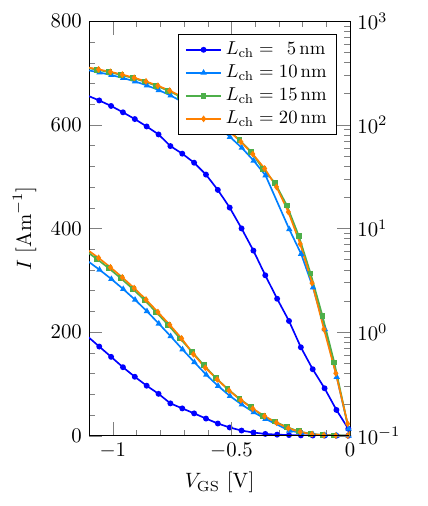}
		\caption{ }
		\label{fig:IVGS_Lch}
	\end{subfigure}\hfil
	\begin{subfigure}[t]{0.32\textwidth}
		\centering
		\includegraphics[height=4.6cm]{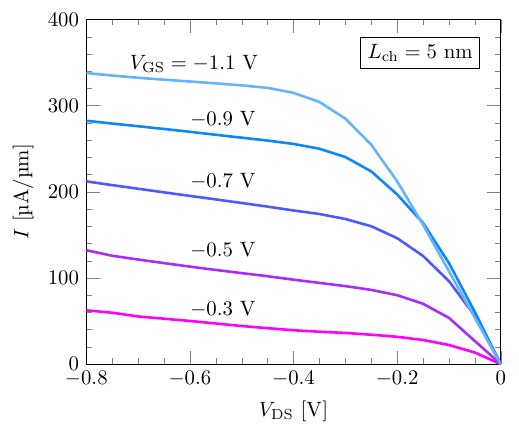}
		\caption{ }
		\label{fig:IVDS_5nm}
	\end{subfigure}\hfil
	\begin{subfigure}[t]{0.32\textwidth}
		\centering
		\includegraphics[height=4.6cm]{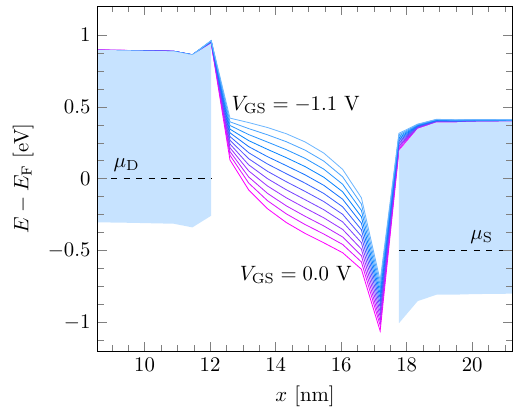}
		\caption{ }
		\label{fig:bands_lhfet_5nm}
	\end{subfigure}
    \caption{(a) Transistor transfer characteristics for different channel lengths $L_{\rm ch} = 5, 10, 15, 20 \  \si{\nano\meter}$. (b) Output characteristics for $L_{\rm ch} = 5 \  \si{\nano\meter}$.  (c)  Band diagram at $L_{\rm ch} = 5 \  \si{\nano\meter}$ for several values of $V_{\rm GS}$, at $V_{\rm DS}=-0.5\ \si{\volt}$.}
    \label{fig:panel_5nm}
\end{figure*}

In~\cref{fig:IVGS_Lch} we explore the scaling behavior of the transistor.
The transfer characteristics are almost overlapping for $L_{\rm G}>10$ nm, while significant performance degradation is observed for $L_{\rm G}=5$ nm.
In this case, the subthreshold swing increases to $\sim\!\!250$ mV/dec and the $I_{\rm ON}/I_{\rm OFF}$ ratio correspondingly drops to 740 and 300 for $V_{\rm DS}=-0.8$ V and $V_{\rm DS}=-0.6$ V, respectively (see~\cref{tab:ionioff}).
Also, the output resistance undergoes a large decrease, settling to $\sim\!\! 12$ k$\Omega\cdot\mu$m at $V_{\rm DS}=V_{\rm GS}=-0.6$ V (see~\cref{fig:IVDS_5nm}).
The performance drop can be traced back to a strongly reduced capability of the gates to modulate the channel potential, as illustrated in panel~\cref{fig:bands_lhfet_5nm}.
Quantitatively, for $L_{\rm ch}=5$ nm the gate efficiency reduces to $0.57$. 
Another marker of performance degradation as the channel length shrinks, is the threshold voltage roll-off $\Delta V_T$ with respect to the $L_{\rm ch}=20$ nm case, which is reported in~\cref{tab:dibl} for the other considered values of $L_{\rm ch}$.

 \begin{table}[!ht]
        \centering
        \caption{$I_{\rm ON}/I_{\rm OFF}$ at $V_{\rm DS}=-0.6$ and $-0.8$ V ratio for different channel lengths.}
        \label{tab:ionioff}
        \begin{tabular}{c  c  c}
        \toprule
            $V_{\rm DS}$ [V] & $L_{\rm ch}$ [nm] &  $I_{\rm ON}/I_{\rm OFF}$ \\
            \hline
            $-0.6$ &  5 &  300\\
            $-0.6$ & 10 & 1180\\
            $-0.6$ & 15 & 1280\\
            $-0.6$ & 20 & 1290\\
            \hline
            $-0.8$ &  5 &  740\\
            $-0.8$ & 10 & 2130\\
            $-0.8$ & 15 & 2300\\
            $-0.8$ & 20 & 2330\\
        \bottomrule
        \end{tabular}
    \end{table}
\begin{table}[H]
    \centering
    \caption{Threshold voltage roll-off $\Delta V_T$ for different channel lengths.}
    \label{tab:dibl}
    \begin{tabular}{ c  c  c }
    \toprule
        $L_{\rm ch}$ [$\si{\nano\meter}$] & $\Delta V_T$ [$\si{\volt}$] & $\Delta V_T$ [$\si{\volt}$]\\
        &  {\small @ $V_{\rm DS} = -0.6 \ \si{\volt}$} &  {\small @ $V_{\rm DS} = -0.8 \ \si{\volt}$}\\
        \hline
         5 & 0.16 & 0.25  \\
        10 & 0.02 & 0.05\\
        15 & 0.01 & 0.03 \\
    \bottomrule
    \end{tabular}
\end{table}
The effect of gate overlaps and underlaps on the $I_{\rm ON}/I_{\rm OFF}$ ratio for $L_{\rm ch}=15$ nm is reported in ~\cref{tab:ion_ioff_overlap}.
As one expects, the presence of overlaps has the effect of improving the $I_{\rm ON}/I_{\rm OFF}$ ratio, as it results in a better electrostatic control over the channel potential; conversely, the presence of underlaps is detrimental to the device performance.
More importantly, a comparison between \cref{tab:ionioff} and \cref{tab:ion_ioff_overlap} highlights that the impact of the gate overlap (underlap) is much stronger than an equivalent gate extension (reduction) with $\delta=0$.
For instance, an increase of $L_{\rm G}$ from 10 nm to 15 nm by symmetrically including 2.5 nm-long overlaps on the source and drain sides, results in a 30\% improvement of the $I_{\rm ON}/I_{\rm OFF}$ ratio, while an equivalent extension of $L_{\rm G}$ with $\delta=0$ only results in a 8\% improvement. 
Analogously, a decrease of $L_{\rm G}$ from 15 nm to 10 nm by symmetrically including 2.5 nm-long underlaps on the source and drain sides degrades the $I_{\rm ON}/I_{\rm OFF}$ ratio by 48\%, while an equivalent gate shrinking with $\delta=0$ only yields a 8\% reduction.
This is a consequence of the modulating effect of the gate over the Schottky barrier thickness already highlighted in \cref{fig:dos_bande_barriera_20nm}.
When $V_{\rm GS}$ decreases, more holes are induced in the {\mos} channel,  leading to a reduction in the size of the depletion regions associated with the Schottky barriers; this results in a thinning effect on the barriers themselves.
In the presence of gate overlaps, the gate is more effective in inducing charge close the {\nbs/\mos} interfaces and therefore in thinning the barriers and modulating the current.
\begin{table}[!ht]
    \centering
    \caption{ $I_{\rm ON}/I_{\rm OFF}$ for $L_{\rm ch} = \SI{15}{\nano\meter}$ at different gate overlap/underlap.}
    \label{tab:ion_ioff_overlap}
    \begin{tabular}{c  c  c}
    \toprule
        $V_{\rm DS}$ [V] & $\delta$ [nm] &  $I_{\rm ON}/I_{\rm OFF}$ \\
        \hline
        $-$0.6 & $-$2.5 &  670\\
        $-$0.6 & \phantom{$-$}0.0 & 1280\\
        $-$0.6 & \phantom{$-$}2.5 & 1540\\
        \hline
        $-$0.8 & $-$2.5 & 1299\\
        $-$0.8 & \phantom{$-$}0.0 & 2300\\
        $-$0.8 & \phantom{$-$}2.5 & 2630\\
        \bottomrule
    \end{tabular}
\end{table}
\begin{figure*}
	\centering
	\begin{subfigure}[t]{0.32\textwidth}
		\centering
		\includegraphics[height=4.6cm]{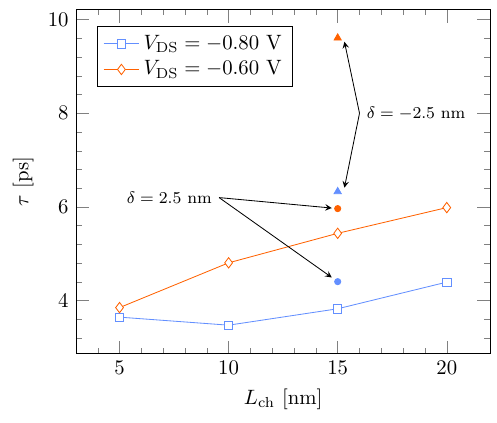}
		\caption{ }
		\label{fig:tau}
	\end{subfigure}\hfil
	\begin{subfigure}[t]{0.32\textwidth}
		\centering
		\includegraphics[height=4.6cm]{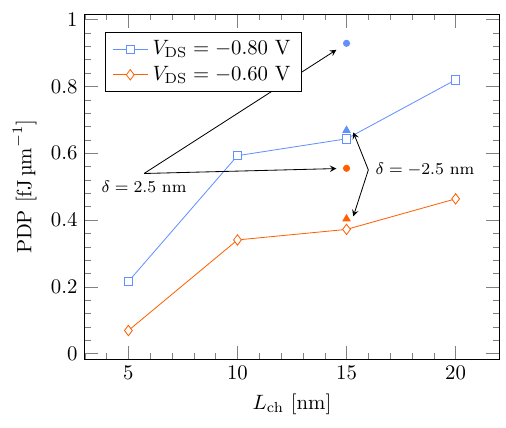}
		\caption{ }
		\label{fig:PDP}
	\end{subfigure}\hfil
	\begin{subfigure}[t]{0.32\textwidth}
		\centering
		\includegraphics[height=4.6cm]{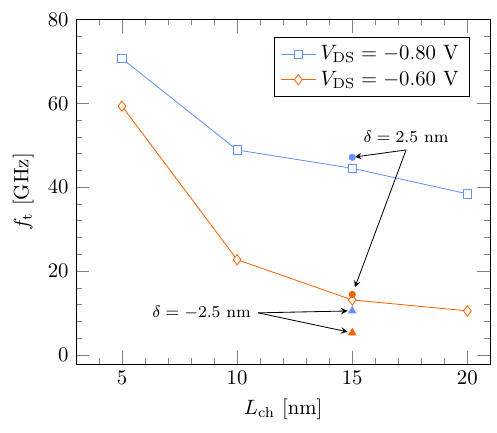}
		\caption{ }
		\label{fig:ft}
	\end{subfigure}
    \caption{Dynamic figures of merit of the transistor as a function of the channel length for $V_{\rm DS}=-0.6$ V and $-0.8$ V. (a) Intrinsic delay time. (b) Power-delay product. (c) Cut-off frequency. In all the cases, the results for a gate overlap (underlap) with $\delta = 2.5\si{\nano\meter}$ ($\delta = -2.5\si{\nano\meter}$) at $L_{\rm ch} = 15 \  \si{\nano\meter}$ are also shown.}
    \label{fig:FOM}
\end{figure*}
Finally, in order to comprehensively evaluate the potential of the device, we analyze several dynamic performance metrics that are relevant to a wide range of technological applications.
These metrics include the intrinsic delay time ($\tau$), the power-delay product (${\rm PDP}$) and the cut-off frequency ($f_{\rm t}$). 
The intrinsic delay time and power-delay product are the basic figures of merit in digital applications, while the cut-off frequency characterizes the device with respect to radiofrequency applications. 

$\tau$ and ${\rm PDP}$ are calculated in a quasi-static approximation using the equations~\cite{TFTfab}
\begin{equation}
    \tau = \frac{Q_{\rm ON} - Q_{\rm OFF}}{I_{\rm ON}},
    \label{eq:tau}
\end{equation}
\begin{equation}
    {\rm PDP} = (Q_{\rm ON} - Q_{\rm OFF}) |V_{\rm DS}|,
    \label{eq:PDP}
\end{equation}
where $Q_{\rm ON}$ and $Q_{\rm OFF}$ are the total charge in the channel in the ON and OFF states, respectively. 
The cut-off frequency $f_{\rm t}$ is evaluated as~\cite{TFTfab}
\begin{equation}
    f_{\rm t} = \frac{1}{2\pi} \frac{g_{\rm m}}{C_{\rm G}} \bigg\rvert_{\max(g_{\rm m})}
    \label{eq:freqT}
\end{equation}
where $C_{\rm G} =\partial Q_{\rm ch}/\partial V_{\rm G}$ is the total capacitance seen from the gates and $g_{\rm m} =\partial I/\partial V_{\rm G}$ is the transistor transconductance.
The results are shown in~\cref{fig:FOM} as a function of the channel length $L_{\rm ch}$.

The intrinsic delay time as a function of $L_{\rm ch}$ is shown in~\cref{fig:tau}.
For $L_{\rm ch}>5$ nm and $\delta=0$, the intrinsic delay time decreases approximately linearly with $L_{\rm G}$.
Larger deviations from this behavior are observed for $L_{\rm ch}=5$ nm, as the values of $\tau$ for $V_{\rm DS}=-0.6$ and $-0.8$ V approximately converge toward a common value of 3.8 ps.
The dependence on $\delta$ is quantified for $L_{\rm ch}=15$ nm.
A large increase of $\tau$ with respect to the case $\delta=0$ is observed for $\delta=-2.5$, due to the previously discussed $I_{\rm ON}$ degradation. 
\newline
The power-delay product shown in~\cref{fig:PDP} increases with increasing $V_{\rm DS}$ and $L_{\rm ch}$, according to~\Cref{eq:PDP}.
A significant increase is observed in the presence of gate overlaps, which indirectly causes an increases of $Q_{\rm ON}$ by enhancing the coupling between the channel and the contacts.
\newline
At fixed $V_{\rm DS}$, the cutoff frequency shown in~\cref{fig:ft} follows a $1/C_{\rm G}\propto1/L_{\rm ch}$ behavior. 
Increasing $V_{\rm DS}$ entails an increase of $g_{\rm m}$ and, in turn, an increase of $f_{\rm t}$. 
In the presence of gate underlaps, the dependence of the transconductance on $V_{\rm DS}$ becomes much weaker and $f_{\rm t}$ tends to settle to a common value of $\sim 10$ GHz. 

\section{Conclusion}

The possibility to fabricate $p$-type field-effect transistors based on 2D materials is essential to develop a CMOS process using 2D materials as transistor channels.
Following recent experimental demonstrations, we have modeled and numerically simulated a \textit{p}-type transistor based on lateral heterojunctions of {\mos} and {\nbs} monolayers.
The metallic {\nbs} regions act as source and drain extensions, mediating between the {\mos} channel and the 3D metal contacts of source and drain.
Our results indicate that the performance of the device is strongly dependent on the modulation of the Schottky barriers at the {\mos/\nbs} interface as a function of both the gate voltage and the drain-to-source bias.
In this respect, the gate length and the gate alignment with the {\mos/\nbs} interface prove to be critical parameters to optimize the device performance.
These geometrical considerations are expected to have a broader applicability and to describe some general tradeoffs in the design of lateral heterojunction field-effect transistors based on two-dimensional materials.

The scaling behavior shows a strong reduction of the $I_{\rm ON}$/$I_{\rm OFF}$ ratio at channel lengths of 5 nm, while the dynamic figures of merit remain dominated by capacitive effects and generally improve as the channel shrinks.

The performance of the device under study does not meet the last IRDS projections~\cite{IRDS2021}, therefore transistors based on {\mos/\nbs} heterojunctions are not competitive with silicon technology for high-performance logic. An important opportunity can come from 3D integration (i.e., the vertical stacking of planar layers of transistors) because devices based on 2D materials can be fabricated with back-end-of-the-line processes on top of silicon MOSFETs. 3D integration can largely improve the logic function density per unit area, still enabling the use of silicon FETs in circuit locations where the best figures of merit are required. For example, effective non-volatile memories with 2D materials have been demonstrated \cite{maregaACSnano} and do not need to exhibit the typical figures of merit of transistors for high-performance logic circuits. In addition, the interface quality of {\mos}/{\nbs} heterojunctions can be significantly improved with progress in fabrication techniques~\cite{Wang2019_1}, yielding better figures of merit.

\bibliographystyle{IEEEtran}


\end{document}